\documentclass[twocolumn,prl,aps, showpacs]{revtex4}
\usepackage{amssymb,amsmath,amsthm,stmaryrd,mathrsfs,bm,verbatim}

\usepackage{hyperref}
\begin{document}
\newtheorem{defn}{Definition}
\newtheorem{Cal}[defn]{Claim}

\title{Comment on Perturbation theory for Heisenberg Operators}
\author{Guowu Meng}

\address{Department of Mathematics\\
Hong Kong University of Science and  Technology\\
Clear Water Bay, Kowloon, Hong Kong} \email{mameng@ust.hk}
\date{\today}


\maketitle

I would like to comment on a recent paper\cite{F} published in
Phys. Rev. A.  In that paper, the author has tried to establish
the following main conclusion: {\it Both the standard perturbation
series and the expanded Magnus exponential operator lead to the
result of Franson and Donegan \cite{FD} for Heisenberg operators}.
He establishes this conclusion through second order and presents a
concrete example to convince the readers that the conclusion must
be true in all orders. Here, I would like to point out that the
author's conclusion is indeed true in {\it all} orders. To prove
this claim, one just need to observe that the formal perturbation
series for Heisenberg operators is unique. The uniqueness is due
to the fact that the initial value problem for a linear 1st
ordinary differential equation has a unique (formal) solution.

\vskip 10pt Recall that the evolution operator is the unique
solution for
\begin{eqnarray}\label{eqn}
\left\{
\begin{matrix}{dU\over dt}&=&\lambda AU\cr
U(0) &= &I\end{matrix} \right.
\end{eqnarray}
where $I$ is the identity operator, $\lambda$ is an imaginary
parameter $1\over i\hbar$, and $A$ is a hermitian operator (the
Hamiltonian in the Schrodinger picture) and could be time
dependent.

For any hermitian operator $\hat O$ in the quantum system, we let
\fbox{$\tau(\lambda A)(\hat O)\equiv[O,\lambda A]$} and \fbox{
$\rho_r(U)(\hat O)\equiv U^\dag\hat O U$}. (Both $\tau(\lambda A)$
and $\rho_r(U)$ are operators on the space of hermitian operators
of the quantum system.) Recall that (hermitian)operator $\hat O$
in the Schrodinger picture and the corresponding operator $\hat
O_H$ in the Heisenberg picture are related via equation $\hat
O_H=\rho_r(U)(\hat O)$. The initial value problem we mentioned
early is
\begin{eqnarray}\label{eqnr}
\left\{
\begin{matrix}{d\over dt}\rho_r(U)&=& \rho_r(U)\tau(\lambda A)\cr
\rho_r(U(0)) &= &I.\end{matrix} \right.
\end{eqnarray}
and it has the following {\bf unique} formal perturbative
solution:
\begin{eqnarray}\label{sol}
\rho_r(U(t))=I+\sum_{n\ge 1}\int_{\Delta_n(t)}d^nt\,\tau(\lambda
A(t_1))\cdots \tau(\lambda A(t_n)),
\end{eqnarray}
where $\Delta_n(t)\equiv\{(t_1,\ldots,t_n)\in
\mathbb{R}^n\,|\,0\le t_1\le \cdots \le t_n\le t\}$. Consequently,
by using the definition of $\tau(\lambda A)$, we have $\hat
O_H\equiv \rho_r(U(t))(\hat O)$ is equal to
\begin{eqnarray}\label{O}
\hskip -15pt & &\hat O+\sum_{n\ge 1}\int_{\Delta_n(t)}\,d^nt\,
[\cdots [\hat O,\lambda A(t_n)],\ldots,\lambda A(t_1)]\cr \hskip
-3pt&=&\hskip -3pt \hat O+\sum_{n\ge
1}\lambda^n\int_{\Delta_n(t)}\,d^nt\, [\cdots [\hat
O,A(t_n)],\ldots,A(t_1)].
\end{eqnarray}
(Note that, equation (\ref{O}) with $A=H$ and $\lambda={1\over
i\hbar}$, being identical to equation (6) in \cite{FD}, is the
main result of Franson and Donegan.)

In view of the uniqueness of the formal expansion for
$\rho_r(U(t))$, it is now clear that the conclusion expected by F.
M. Fern\'{a}ndez is valid in all orders.

\vskip 10pt {\bf Justification of equations (\ref{eqnr}) and
(\ref{sol})}. To solve equation (\ref{eqnr}), let
\begin{eqnarray}\label{fexp}
\rho_r(U(t))=u_0(t)+\lambda u_1(t)+\cdots=\sum \lambda^n u_n(t)
\end{eqnarray}
be the formal perturbative expansion of $\rho_r(U(t))$ in powers
of $\lambda$. Plugging (\ref{fexp}) into (\ref{eqnr}), we have
$u_0(t)=I$ and a sequence of initial value problems
\begin{eqnarray}\label{eqns}
{d(\lambda u_n)\over dt}&=& u_{n-1}\tau(\lambda A)\cr \lambda
u_n(0) &=& 0
\end{eqnarray}
Solving (\ref{eqns}) inductively, the unique solution in equation
(\ref{sol}) is then obtained. Next, we note that, to arrive at
equation (\ref{eqnr}) we just need to verify from the definitions
that 1) $\rho_r(U(0))(\hat O)=O$, 2)$\left({d\over
dt}\rho_r(U(t))\right)(\hat O)\equiv{d\over dt}\left(U^\dag(t)\hat
O U(t) \right)=\left(\rho_r(U(t))\tau(\lambda A(t))\right) (\hat
O)$. The actual verification is just a simple computation based on
definitions and equation (\ref{eqn}) and is omitted here due to
the restriction of the length of this comment. (For those people
who know representation theory, the quickest way to obtain
equation (\ref{eqnr}) is to apply the right adjoint representation
to equation (\ref{eqn}). In fact, the quickest way to obtain
equation (\ref{O}) is to apply the right adjoint representation to
the formal perturbative expansion formula for solution of equation
(\ref{eqn}) --- a formula in many standard textbook of quantum
mechanics.)

\end{document}